\documentclass[aps,pre,twocolumn,amsmath,amssymb,groupedaddress,showpacs,showkeys]{revtex4-1}
\usepackage{graphicx}
\begin{document}

%\preprint{}

\title{Traffic gridlock on a  honeycomb city}

\author{L.E. Olmos}
\email[]{leolmoss@unal.edu.co}
\affiliation{Simulation of Physical Systems Group, CeiBA-Complejidad, Physics Department, National University of Colombia, Cra. 30 \# 45-03, Ed.404, Of, 348, Bogot\'a D.C., Colombia}
%\thanks{}
\author{J. D. Mu\~noz}
\email[]{jdmunozc@unal.edu.co}
\affiliation{Simulation of Physical Systems Group, CeiBA-Complejidad, Physics Department, National University of Colombia, Cra. 30 \# 45-03, Ed.404, Of, 348, Bogot\'a D.C., Colombia}

%Collaboration name if desired (requires use of superscript address

\date{\today}

\begin{abstract}
As a clear signature of modern urban design concepts, urban street networks in dense populated zones are evolving nowadays towards grid-like layouts with rectangular shapes, and most studies on traffic flow assume street networks as square lattices. However, ideas from forgotten design schools bring unexplored alternatives that might improve traffic flow in many circumstances. Inspired on an old and almost in oblivion urban plan, we report the behavior of the Biham-Middleton-Levine model (BML) \-- a paradigm for studying phase transitions of traffic flow \-- on a hypothetical city with a perfect honeycomb street network.  In contrast with the original BML model on a square lattice, the same model on a honeycomb does not show any anisotropy or intermediate states, but a single continuous phase transition  between free and totally congested flow, a transition that can be completely characterized by the tools of classical percolation. Although the transition occurs at a lower density than for the conventional BML, simple modifications, like randomly stopping the cars with a very small probability or increasing the traffic light periods, drives the model to perform better on honeycomb lattices. As traffic lights and disordered perturbations are inherent to real traffic, these results question the actual role of the square grid-like designs and suggests the honeycombs as an interesting alternative for urban planning in real cities.

\end{abstract}


\maketitle

As cities turn denser, urban networks tend to adopt a squared-lattice shape \cite{barthelemy}, and many traditional urban planning styles, like the one spaniards and portugueses disseminated through all Latin America, are grounded on such square patterns \cite{rama}. Following this trend, most prominent studies on city traffic adopt square lattices\cite{schads1, daganzo2007, manhattan}. Despite modern urban planners claim that this design favors connectivity, the question if a square design optimizes traffic flow has not being studied systematically. In contrast, Nature usually opts for other alternatives. Hexagonal structures in two dimensions are present in cellular tissues\cite{mombach,gibson}, bee honeycombs\cite{honey} and soap bubbles \cite{bubbles,waire}. Such patterns arise by minimizing surface energy on a fixed area \cite{vonNeumann}.  
Inspired by Nature, humans have also implemented hexagonal tesselations in a wide range of disciplines, including  structured materials \cite{fractal, lorna}, wireless networks\cite{phone}, computer graphics\cite{computegraphs}, etc.
 However, in the realm of the urban design, street patterns based upon hexagonal block are just a theoretical alternative which has fallen into oblivion with almost no practical applications (see \cite{ben-joseph} and refs. therein), but hiding possible unexplored  solutions for the overwhelming problem of traffic flow in modern cities. 

The BML model is the simplest traffic cellular automaton able to exhibit self-organization, pattern formation and phase transitions \cite{biham92, tadaki, kertesz, gupta}. Although the model oversimplifies the city, much extensive research has been based on it \cite{lightstrategy, kinetic,nagatani,chowd}.  The original model describes two species of cars (east-running and north-running cars) moving by turns on a two-dimensional square lattice with periodic boundary conditions. Driven by car density, the system falls into three different phases: free flow (all vehicles move), jammed phase (all vehicles are stuck) and intermediate states where jams and free flow  coexist on a wide density range\cite{raissa1,raissa2,wang}. A recent study have shown that such intermediate states are a consequence of the anisotropy inherent to the model\cite{OlmosMunoz2015}, which produces two different phase transitions: one if the system is longer in the flow direction (longitudinal) and other if the system is longer in the perpendicular one (transversal).  It has also been reported that this intermediate phase disappears when some kind of randomization is introduced\cite{ding1,ding2,raissa2}, or the traffic periods for the two cars are increased \cite{trafficlights}. Some other extentions include free boundary conditions \cite{open}, four directions for the cars \cite{fourDirections} or 3D implementations \cite{3D}. In contrast, the role played by the network topology has been overlooked and, there are very few studies considering the BML model on different lattices: square lattice generalizations with extra sites in the bonds\cite{decorated, decorated2} and triangular lattices where three species of cars are considered\cite{triangular1, triangular2}. In all cases a more complex behavior with different jammed phases is observed.

The main goal of this work is to test the BML traffic model\cite{biham92} on honeycomb lattices. The intention is to explore if using a different lattice affects the jamming transition and, eventually, when a honeycomb lattice offers a better performance than the square one. As in the original model, we will implement two car species moving by turns on a lattice with periodic boundary conditions, which can be closed on a torus in three different ways. Surprisingly, all systems show a single well-defined phase transition, although there is still a preferred flux direction and, moreover, there are cases where the BML performs better on honeycomb lattices than on square ones. So, this work questions the assumption that square grids are always optimal and suggests honeycombs as interesting alternatives for urban designers.

\begin{figure}[t]
\includegraphics[width=0.44\textwidth]{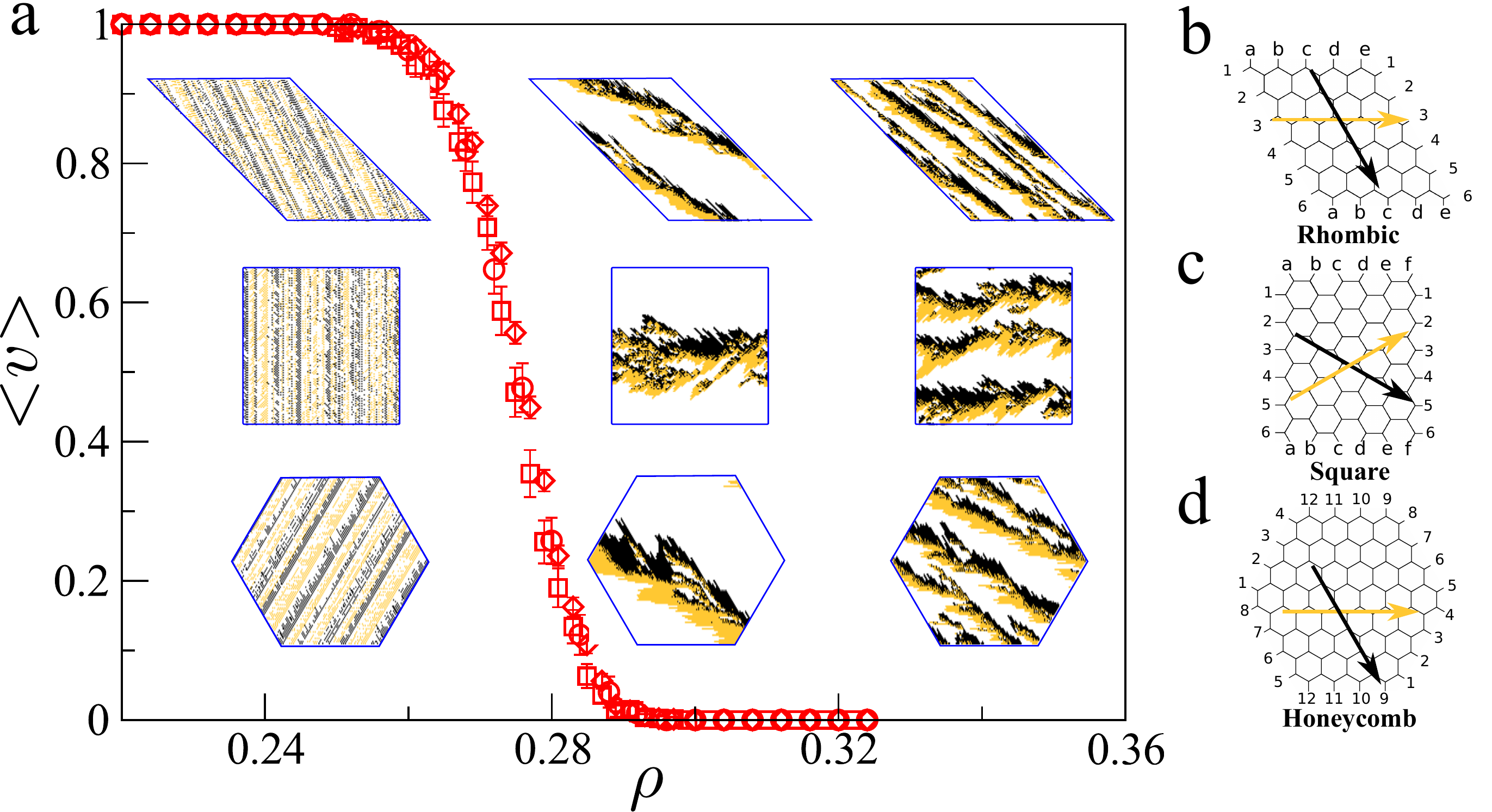}
\caption{(color online) Average velocity $\left< v\right>$ vs density $\rho$ (solid red line) for the BML model on $128\times128$ honeycomb lattices. Insets show snapshots for free flow (left), one global jam (center) and random jams (right) on lattices with three boundaries: rhombic  (diamonds and (b)), square  (squares and (c)) and honeycomb (circles and (c)). The flow direction is defined by just two (yellow and black arrows) of the three reflection symmetry axes.}
\label{transition}
\end{figure}

{\it Model.} 
Consider two types of cars moving zig-zag in two different directions, yellow and black, on a honeycomb-like lattice with periodic conditions (Fig. \ref{transition}). Each node is connected with three others and can be in one of three states: empty, occupied by a yellow car, or occupied by a black one. The cars are initially randomly distributed over the lattice sites with spatial density $\rho$. The fully deterministic dynamics is as follows: On even (odd) steps, all yellow (black) cars attempt to advance one lattice site on his zig-zag pattern. If the site ahead of a car (in color direction) is currently empty, it advances; otherwise, it remains stationary. 
The system is implemented on a torus, i.e. with periodic boundary conditions, as in the original model. Nevertheless, there is no unique way to close an hexagonal lattice on a torus, but three \cite{stojmenovic}: square, rhombic  and honeycomb (Fig \ref{transition}(b-d)). We shall consider all these three tori in the most part of our analysis.

{\it Absence of anisotropy.}
Starting the simulations from random configurations, the system reaches one of its limiting states after a transient period. If the system size is large enough ($L>64$), there are only two different limiting states (Fig. \ref{transition}(a)): a free-flow phase, where all cars move freely every time step ($v$$=$$1$) and a jammed phase, where no cars move ($v$$=$$0$). Contrary to the original model, there are no intermediate states, and the system exhibits a sharp jamming transition between these two phases (Fig \ref{transition}(a)).  

As in the original model, there is a preferred flow direction: the one bisecting the two directions for cars and, in consequence, it could be possible to find a similar anisotropy in the correlation length. Let us start by studying the isotropy of the system. If the density is large enough, the system reaches a jamming state after a transient period. Following the methods applied in \cite{OlmosMunoz2015}, we define the parallel (perpendicular) spatial correlation function \cite{tadaki} as 
\begin{equation}
G_{\parallel (\perp)}(\vec{r'})=\frac{1}{N}\left\langle\sum_{\vec{r}} \sigma(\vec{r})\cdot\sigma(\vec{r}+\vec{r'})\right\rangle \quad,
\end{equation}
where $\sigma(\vec{r})$$=$$1(0)$ if the site with position $\vec{r}$ is occupied(empty), $N$ is the total number of cars and $\vec{r'}$ is a vector in the direction $\parallel$($\perp$) you want to compute the correlation function along. The symbol $\langle\rangle$ denotes averages over final jammed configurations starting from different random initial conditions at densities slightly above the jamming transition.  The correlation functions are fitted with exponentials $G_{\parallel(\perp)}\propto\exp(-r/\xi_{\parallel(\perp)})$ to estimate the correlation lengths $\xi_{\parallel(\perp)}$ in each direction.  The anisotropy exponent $\theta$ can be estimated numerically from the fact that, close to the critical point, the two correlations lengths must be related by $\xi_\parallel\sim\xi_\perp^{\theta}$\cite{binder, redner}. 

\begin{figure}[t]
\includegraphics[width=0.33\textwidth]{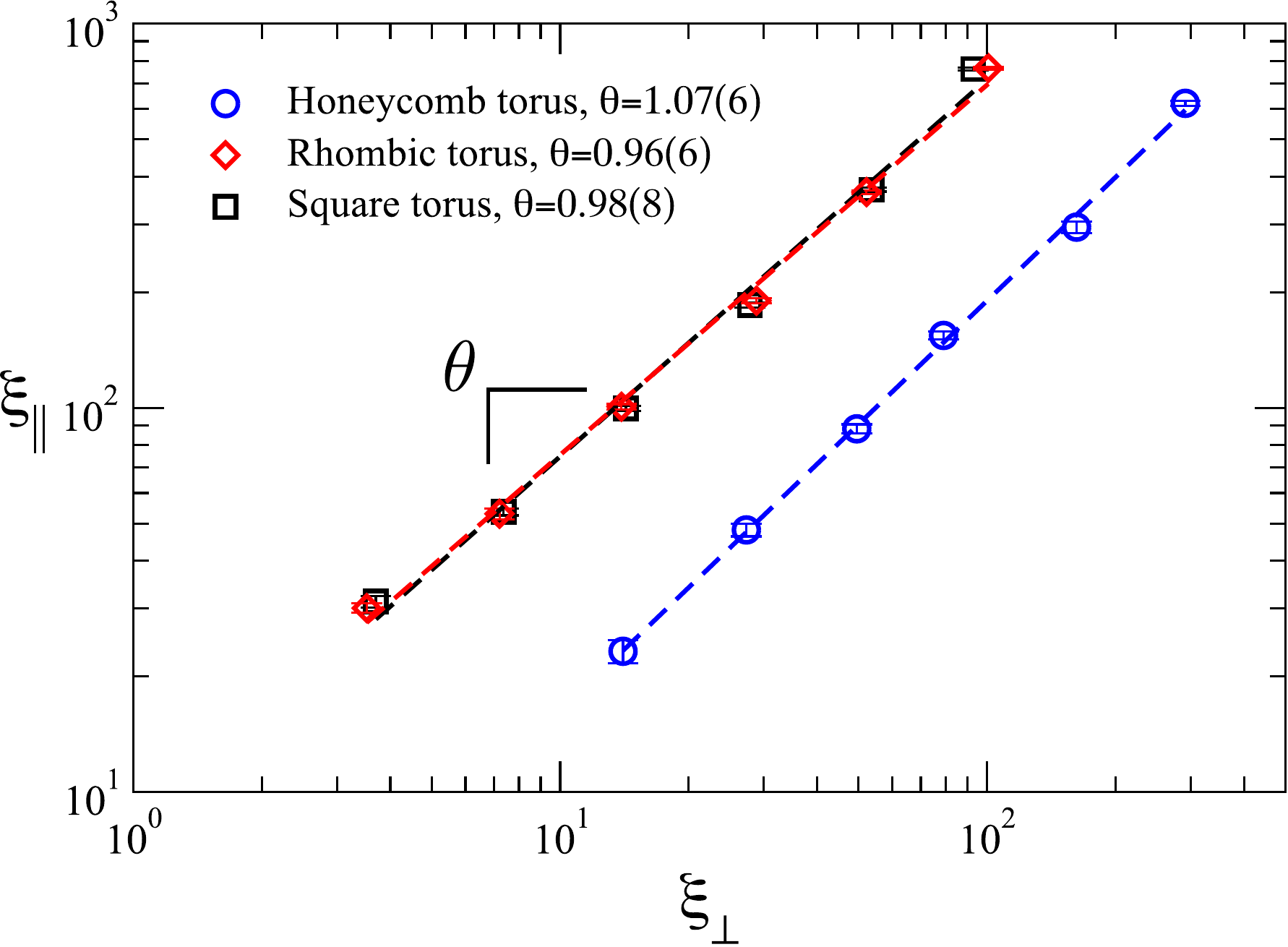}
\caption{(color online) Longitudinal $\xi_\parallel$ and transversal $\xi_\perp$ correlation lengths from final configurations at densities $\rho$ in the range $[0.265-0.310]$ for honeycomb lattices  of different sizes with the three boundary conditions. Each point is an average over 50 configurations. The dashed lines show the power-law fits with anisotropy exponents $\theta$$\approx$$1.0$, i.e. the system behaves isotropic. Here and everywhere the error bars are 3$\sigma$.}
\label{Corr}
\end{figure}

Figure \ref{Corr} presents the correlation lengths computed from final configurations of the BML model for the three different honeycomb tori with different sizes and at densities close to the threshold transition. A power-law fit gives values for $\theta$ very close to 1, meaning that the system can be considered isotropic, such that the standard finite-size scaling (FSS) theory is suitable for describing the phase transition. Indeed, simulations on systems with different aspect ratios (not shown here) show no difference on the transition. This surprising result is, therefore, not a consequence of the preferred flow direction alone, but also of the grid itself.

\begin{figure*}[]
   \includegraphics[width=0.8\textwidth]{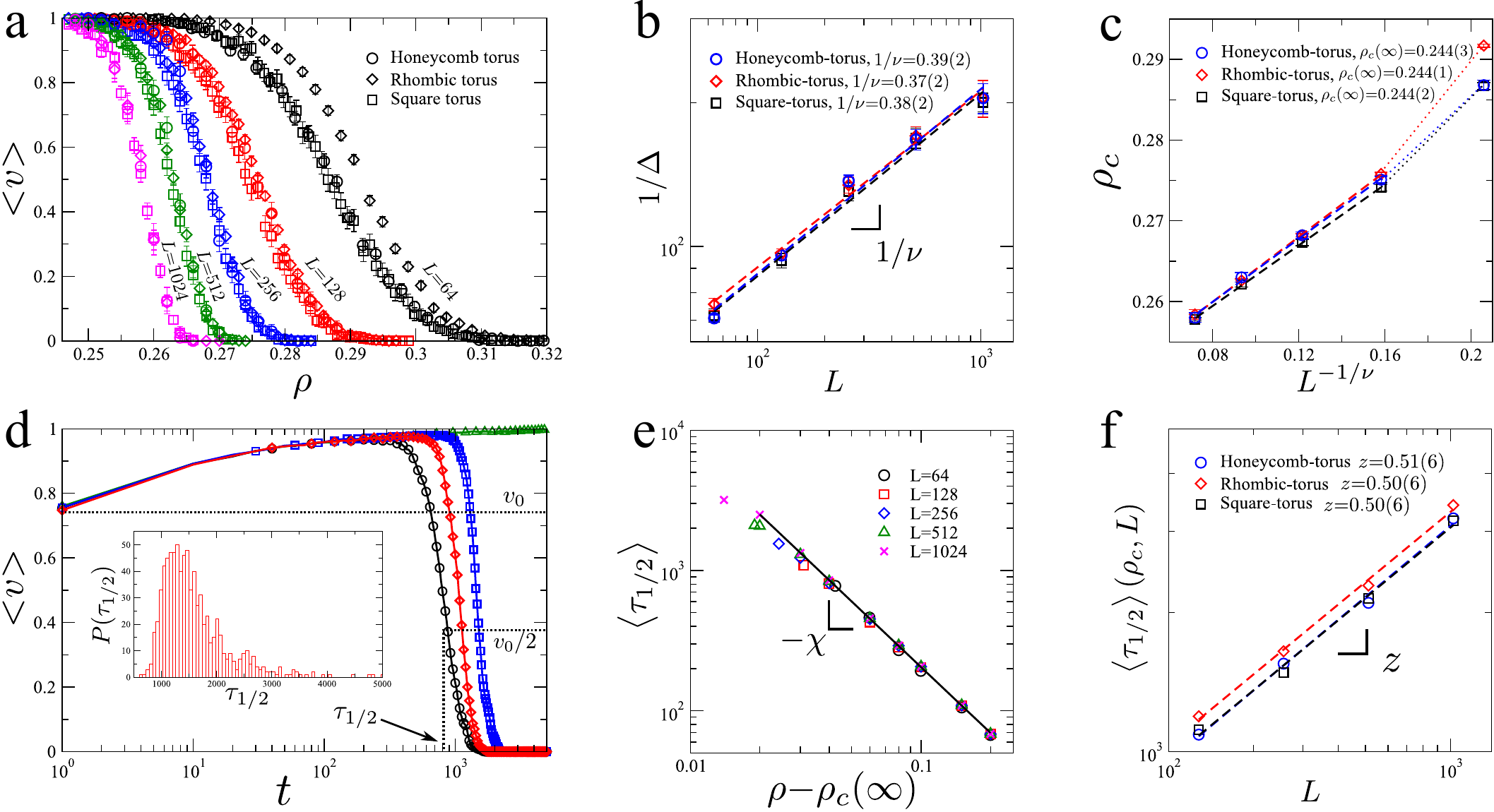}
\caption{(color online) Finite-size scaling analysis for the dynamical phase transition. (a)  Transition curves for the three types of torus (symbols) with five different system sizes (colors), ranging from $L$$=$$64$ to $L$$=$$1024$.  Each point is averaged over 2000 (1000)  final configurations for $L$$\le$$512$($L$$=$$1024$), obtained after convergence ($v$$=$$0$ or $v$$=$$1$) or after $2 \times10^5$ time steps (whichever comes first). (b) Scaling of the transition width $\Delta (L)$. Dashed lines are  power-law fits for the three tori, giving $1/\nu$$=$$0.38(3)$ on average. (c) Scaling of the finite critical density. Because of strong finite-size effects, we neglect $L$$=$$64$, and obtain $\rho_{c}(\infty)$$=$$0.244(3)$ on average. (d) Average speed in function of time for five configurations. Dotted lines show the definition of the relaxation time $\tau_{1/2}$. The inset evidences that $\tau_{1/2}$ follows a lognormal distribution. (e) Mean relaxation time  $\tau_{1/2}$ for densities above $\rho_{c}(\infty)$ on the honeycomb-torus (results on other tori are quite similar). The slope gives on average a critical exponent $\chi$$=$$1.55(2)$. (f) Scaling of the relaxation time at the critical point  $\tau_{1/2}(\rho_c)$. On average, we obtain a dynamical critical exponent $z$$=$$0.50(6)$. Each point on the last two figures is averaged over 100 configurations.}
\label{allfss}
\end{figure*}

{\it The jamming transition.}
Fig. \ref{allfss}(a) shows the transition curves for several system's sizes, ranging from $L$$=$$64$ to $L$$=$$1024$. In the honeycomb-torus case, the size $L$ denotes the torus with the number of nodes closest to $L^2$\footnote{A honeycomb torus of size $n$ has $6n^2$ nodes and $n$ hexagons between the center and boundary. Thus, a $L^2$  torus  actually corresponds to a torus in which $n$ is the closest whole number of $L/\sqrt{6}$}. As in many models with phase transitions in statistical physics (e.g. percolation \cite{StaufferAharony}), the value of of the critical density $\rho_c$ decreases with system size, reaching a critical value $\rho_c$ as the system size approaches infinity.  By fitting the transition curves with an error function, figures \ref{allfss}(b) and (c) show that the transition width and the density threshold scale as  \cite{Torquato}
\begin{equation}
\Delta(L)\sim L^{-\frac{1}{\nu}} \text{   and   } \left|\rho_c - \left< \rho_c(L)\right>\right| \sim L^{-\frac{1}{\nu}} \quad.
\label{scale}
\end{equation}
The values obtained for $\nu$ and $\rho_c(\infty)$ are very similar for the three tori. On average, we obtain $1/\nu$$=$$0.38(3)$ and $\rho_c(\infty)$$=$$0.244(3)$. 

To investigate the dynamics of the model in the jammed state, let us define $\tau_{1/2}$\cite{tadaki} as the time when the average speed is half of the initial speed (Fig. \ref{allfss}(d)). This relaxation time follows a lognormal distribution and, therefore, its mean value can be estimated as $\left<\tau_{1/2}\right>=\exp(\mu+\sigma^2/2)$, with $\mu\simeq\frac{1}{n}  \sum_k \ln{\tau_{1/2}}_k $ and $\sigma^2\simeq\frac{1}{n}  \sum_k (\ln{\tau_{1/2}}_k-\mu)^2$ .

In the jammed phase ($\rho$$>$$\rho_c$),  Fig. \ref{allfss}(e) shows that $\left<\tau_{1/2}\right>$ is independent of the system size and scales as $\left<\tau_{1/2}\right>\sim(\rho-\rho_{c})^{-\chi}$  with $\chi$$=$$1.55(2)$. In addition, the values of $\tau_{1/2}$ at the critical density $\rho_c$ scales with system size as $\left<\tau_{1/2}\right>(\rho_{c},L)\sim L^{z}$, with  $z$$=$$0.50(6)$ (Fig. \ref{allfss}f). The finite size scaling theory suggests that above the transition point $\chi/\nu$$=$$z$$=$$0.56(5)$, in fair agreement with the value above.

{\it A mean-field analysis.}
Interestingly, the critical density  can be approximated by using a {\it naive} mean-field analysis, inspired by \cite{wangimproved}. Consider the mean velocity of yellow cars (by symmetry, the reasoning is also valid for black cars). A yellow car will stop either because it is blocked by a black car or by another yellow car. On honeycomb lattices, there is almost no difference between these two types of interactions. At a random initial configuration, the probability that a car is blocked is $\rho$, that is, at the beginning of the simulation the proportion of stopped cars must be equal to $\rho$.  Since black (yellow) cars spend on average a time $1/v$ on a site, they will reduce the speed of yellow cars from unity by $\rho/v$. Hence, a self-consistency equation for the average speed $v$ will be 
\begin{equation}\label{meanfield}
v=1-\frac{\rho}{v}\quad,
\end{equation}
which gives $\rho_c$ as the critical density at which the equation ceases to give a real solution. That occurs at $\rho_c$$=$$0.25$, very close to the value of $0.244(3)$ obtained from finite size scaling. 

{\it A comparison with the square lattice.} \begin{figure*}
\centering
\includegraphics[width=0.8\textwidth]{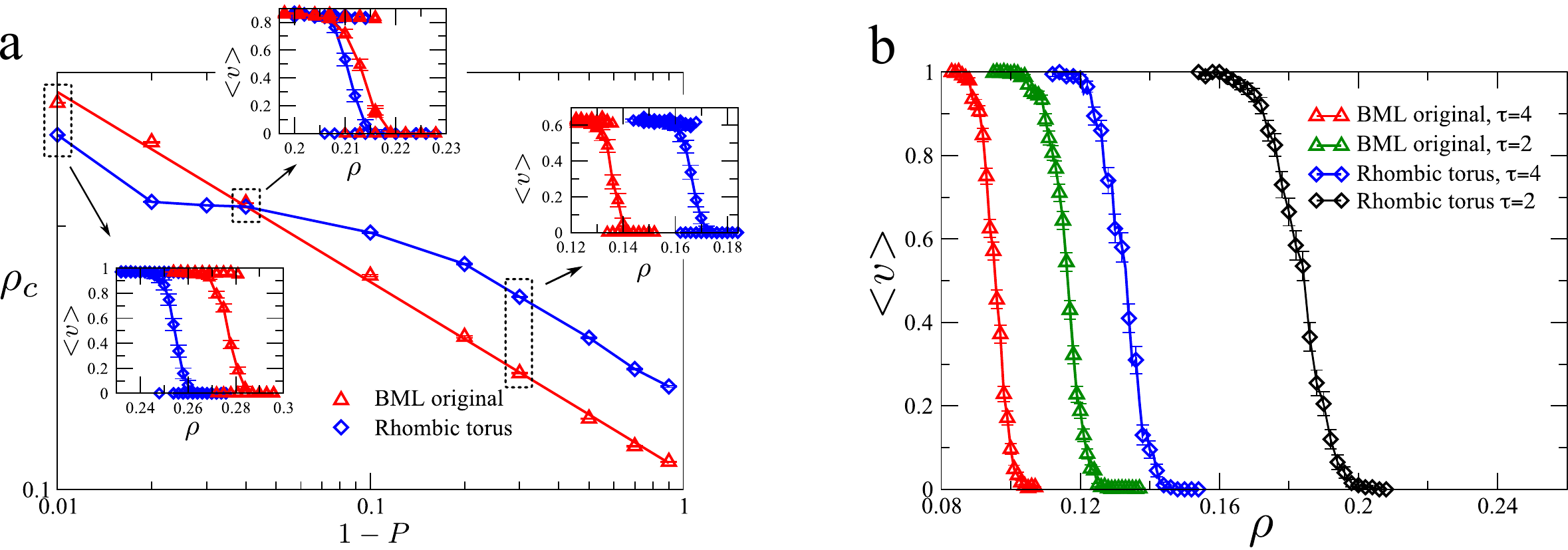}
\caption{(color online) Effects of two modifications of the BML model on both rhombic tori (diamonds) and   square lattices (triangles). (a) Effect of including a random update, where cars move with probability $P$ if the target site is empty. The figure shows the critical density $\rho_c$ as a function of $1-P$ for lattice sizes $L$$=$$128$. Insets show the transition curves for three values of $P$. (b) Effect of increasing the traffic light period $\tau$. The figure shows transition curves for $\tau$$=$$2$ and $\tau$$=$$4$ on lattices with size $L$$=$$256$.  Each point in both figures is  averaged on 400 runs. Measurements are obtained after  $6 \times10^5$ time steps or until convergence (whichever comes first).}\label{random}
\end{figure*}
The critical density $\rho_c$$=$$0.244(3)$ for the BML model on a honeycomb is lower that the value of $0.283(2)$ for the lowest transition on a square lattice\cite{OlmosMunoz2015}. However, this order is reversed in at least two cases. First, let us remove full synchrony by introducing a random update  \cite{ding1}, where a car advances with probability $P$$<$$1$ if the target site is empty; a modification that also destroys the intermediate state in of the BML model on square lattices \cite{ding1,ding2,raissa2}.  Figure \ref{random}a compares the critical density of the model as function of $1-P$ on a rhombic torus with the one on a square lattice. The BML on a square lattice follows a power law behavior, with $\rho_c$$\propto$$(1-P)^{-0.22(1)}$, behaves better only for a narrow interval. Below $P$$=$$0.96$, the honeycomb lattice overcomes the square one and behaves better, that is with a higher critical density.
Second, we have also studied the effect of increasing the traffic-light periods, that is cars on each direction have the chance to advance in $\tau$ consecutive time steps ($\tau$$=$$1$ for the original model). This also destroys the intermediate states on the original BML model and, furthermore, produces a spatial phase separation with small global speeds at intermediate densities \cite{trafficlights}. Again, rhombic tori show higher critical densities than square lattices, even for $\tau$$=$$2$ or $\tau$$=$$4$ (Figure \ref{random}b). These results suggest that the model on a honeycomb is more resilient against small perturbations than on a square lattice.\\

{\it Conclusions and discussions.} 
We have shown that  the BML model with two flow directions behaves isotropically on honeycomb networks. There are no intermediate states, and a sharp transition from the moving phase to the jamming phase is observed at a critical vehicle density. Despite the fact that there is a preferred flow direction, the correlation length shows to be isotropic,. This surprising result may be a consequence of the symmetries of the honeycomb. Indeed, it has been shown that high-order tensors on a hexagonal lattice (the dual lattice of a honeycomb) are isotropic up to second order in the grid size\cite{wolfram}. If this is the reason for such isotropy or not will be an interesting subject of future research.

By performing a classical scaling analysis, we characterized completely the transition, measuring the critical density and three critical exponents. Although the model shows a lower critical density than on square lattices, this issue is reversed by introducing small and simple perturbations, like increasing the traffic light periods or including a random update with very low probabilities to brake. Street patterns based upon hexagonal blocks were proposed by several planners in the early 20th century \cite{ben-joseph}. Despite urban designers demonstrated the economic advantages and efficient land use of hexagonal plans, this idea never ceased to be a theoretical alternative to the rectangular grid, never implemented in urban street patterns.  Furthermore, the contemporary movements of New Urbanism claims that square grid layouts increase the connectivity\footnote{see http://www.newurbanism.org/}, dispersing traffic and reducing driving times, because they are assumed to be mixed-use, walkable, and more pedestrian friendly. However, such assumptions are criticized by practical considerations \cite{ben-joseph}. Indeed, empirical data about safety \cite{accidents,accidents2} suggest that 4-legs intersections, ubiquitous in square grids, increase both the number of crashes and injuries significantly, suggesting to reconsider urban layouts where T-junctions predominates ({\it cul de sac}, radburn, fused grid). Moreover, city planners use to restrict flow direction emulating  T-junctions. Honeycomb grids emerge as an unifying idea.

Our results suggest that the BML model on hexagons under perturbations is more robust than on squares. As the perturbations included, i.e. traffic lights and disorder, are crucial in real traffic, this work questions the real role of the square grid-like designs and supports honeycombs as an interesting alternative for urban  densification processes.  

\begin{acknowledgments}
We thank the Centro de Estudios Interdisciplinarios B\'asicos y Aplicados en Complejidad, CeiBA-Complejidad, and the Universidad Nacional de Colombia for financial support. We are indebted M. C. Gonzalez for hospitality and useful discussions.
\end{acknowledgments}

\bibliography{apssamp}

\end{document}